\documentclass{jaa}
\usepackage{natbib}
\usepackage{graphicx}

\begin{document}\sloppy

\title{Revisiting the old end of the Milky Way open cluster age function}


\author{Andr\'es E. Piatti\textsuperscript{1,2,*}}
\affilOne{\textsuperscript{1}Instituto Interdisciplinario de Ciencias B\'asicas (ICB), 
CONICET-UNCUYO, Padre J. Contreras 1300, M5502JMA, Mendoza, Argentina; \\
\textsuperscript{2}Consejo Nacional de Investigaciones Cient\'{\i}ficas y T\'ecnicas 
(CONICET), Godoy Cruz 2290, C1425FQB,  Buenos Aires, Argentina}


\twocolumn[{

\maketitle

\corres{andres.piatti@fcen.uncu.edu.ar}


\begin{abstract}
The age distribution of the open cluster system is a key piece
of information to decipher the star formation history of the Milky Way disk.
Recently, a remarkable earlier drop of its older end was found, which caught our attention. 
Precisely, we analyzed in detail
the population of open clusters older than 1  Gyr located inside a circle of 2.0 kpc from the Sun
contained in the Milky Way Star Cluster catalog, using the Data Release 3.0 of
the {\it Gaia} survey, and found that it contains a slightly larger old
open cluster population with respect to that witnessing the earlier drop age
distribution. However, there are  still some aspects that deserve further  attention in order to
undoubtedly handle a statistically complete cluster sample, that allows us to
comprehensively know the older end of the open cluster age distribution function.
We discuss some reasons that
affect such a completeness, among them, the photometric depth of the database employed,
the performance of machine learning techniques used to recognize open clusters, the
cleaning of cluster color-magnitude diagrams from field star contamination, etc.
\end{abstract}

\keywords{Galaxy: open clusters and associations: general --- methods: data analysis --- methods: statistical}

}]



\def\msun{\hbox{M$_\odot$}}

\def\reference{In: Pei, J., Tseng, V.S., Cao, L., Motoda, H., Xu, G. (eds) Advances in Knowledge
Discovery and Data Mining. Lecture Notes in Computer Science(), vol 7819. Springer, Berlin,
Heidelberg.}

\def\mnras{MNRAS}
\def\aap{A\&A}
\def\apjs{ApJS}
\def\apj{ApJ}
\def\aj{AJ}
\def\araa{ARAA}
\def\na{New A.}

\section{Introduction}

Recently, \citet{andersetal2021} showed that the population of Milky Way
open clusters older than $\sim$ 1 Gyr, located closer than 2 kpc from the Sun,
is much smaller than previously known, which brings important implications to 
our comprehensive knowledge of the formation and evolution of the Milky Way
open cluster system. They arrived at this result from the 
comparison of their derived open cluster age function with that obtained from 
the Milky Way Star Cluster catalog 
\citep[MWSC;][]{kharchenkoetal2013,piskunovetal2018,krumholzetal2019}.
The decrease in the number of open clusters older than $\sim$ 1 Gyr in the
\citet{andersetal2021}'s sample comes from the fact that they could not
confirm as genuine open clusters many of those included in the MWSC 
catalog in that age range.  \citet{andersetal2021}'s compilation 
includes 268 open clusters older than 1 Gyr (also found in the MWSC catalog); 
an amount nearly twice the number of old open clusters in the
MWSC catalog not detected by them. The fundamental parameters
(age, distance, metallicity, etc) of the latter have been extensively used 
in studies of the open cluster system and the Milky
way disk \citep[e.g.][]{joshietal2016,kharchenkoetal2016,dibetal2018}, 
which means that they have been considered real physical systems.

As mentioned by \citet[][]{cantatgaudinetal2018},
on which \citet{andersetal2021} based their work, there are some reasons that 
explain the lack of detection of open clusters in
their analysis, among them, the density of the background, the
interstellar extinction, the cluster's star richness, the difference
of proper motion between the cluster and the field, the cluster age, etc. 
Indeed, older open clusters, and particularly those more distant, contain
Main Sequence stars fainter than the magnitude limit used for detecting
open clusters in the \citet{andersetal2021}'s sample. Therefore, it arises
necessary to revisit those non-detected clusters in
the MWSC catalog with the aim of examining their physical nature.

In contrast with this resulting decrease in the number of confirmed old open clusters,
evidence of enhanced cluster formation episodes
with two primary excesses at $\sim$ 10-15 Myr and 1.5 Gyr were found
by \citet{p10}. He used 1787 open clusters from the \citet{detal02}'s
catalog, and confirmed both age peaks when restricting the
sample of open clusters to those located in the solar neighborhood, with 
the aim of avoiding incompleteness effects. Moreover, recent search
for new open clusters based on the {\it Gaia} DR3 database 
\citep{gaiaetal2016,gaiaetal2022b} and a variety of machine learning techniques
have found many new open clusters older than 1 Gyr 
\citep[see, for instance,][among others]{kounkeletal2020,castroginardetal2022,haoetal2022,qinetal2023}.
In this sense, previous efforts in assessing the physical reality of open 
clusters showed the importance of dealing with a statistically complete sample of 
genuine open clusters for studies of the recovery of the open cluster formation 
history and their destruction rate, the structure of the Milky Way disk, etc 
\citep{pietal11b,p17c,diasetal2021w}.

Precisely, the main goal of this paper is to reanalyze the sample of open clusters
older than 1 Gyr included in the MWSC catalog that are not present 
in the open cluster age function obtained by \citet{andersetal2021}, with
the aim of providing a robust assessment on their physical nature.
We describe the analysis strategy in Section 2 and discuss the derived results
in Section 3. We include in the Appendix all the supporting material
in order to ease the reading of the text.

\begin{table*}
\caption{Literature search for selected MWSC catalog's open clusters.\\
Open clusters are listed according to their literature's ages in descendent\\
order.}\label{tab1}
\begin{tabular}{lccc}
\topline
Name        & log(age /yr) &  Ref. &  log(age /yr)$_{\rm MWSC}$\\\hline
\multicolumn{4}{c}{Open clusters older than 1 Gyr}\\\hline
NGC~1663            & 9.70$\pm$0.20 & 4 & 9.40 \\
FSR~0070            & $\approx$ 9.70 & 5 & 9.57 \\
ESO~447-29          & 9.60$\pm$0.25 & 2 & 9.50 \\
ESO~522-05          & 9.50$\pm$0.20 & 3 & 9.50 \\
ESO~137-23          & 9.45$\pm$0.15 & 1 & 9.55 \\
vdBergh-Hagen~118   & 9.45$\pm$0.05 & 8 & 9.05 \\
NGC~7036            & 9.35$\pm$0.10 & 4 & 9.45 \\
Koposov~77          & 9.10$\pm$0.10 & 6 & 9.45 \\
ESO~425-15          & 9.05$\pm$0.15 & 4 & 9.01 \\
Bica~6              & 9.00$\pm$0.05 & 7 & 9.19 \\\hline
\multicolumn{4}{c}{Open clusters younger than 1 Gyr}\\\hline
ESO~429-13          & 8.95$\pm$0.15 & 1 & 9.02 \\
ASCC~92		& 8.85$\pm$0.05 & 12 &  9.08 \\
Collinder~351       & 8.85$\pm$0.10 & 9 & 9.00 \\
ESO~426-26          & 8.80$\pm$0.10 & 4 & 9.13 \\
Lyng\aa~8           & 8.60$\pm$0.10 & 4 & 9.29 \\
FSR~0224            & 6.74$\pm$0.05 &10 & 9.01 \\\hline 
\multicolumn{4}{c}{Not confirmed open clusters}\\\hline
ESO~436-02          & --            & 11 & 9.1 \\\hline
\end{tabular}
\tablenotes{\noindent Ref: 1) \citet{piattietal2019d}; 2) \citet{monteiroetal2017}; 
3) \citet{tadross2008}; \\ 4) \citet{angeloetal2019a}; 5) \citet{bicaetal2008c};
6) \citet{yadavetal2011}; \\ 7) \citet{bonattoetal2008}; 8) \citet{p16b}; 
9) \citet{angeloetal2020}; \\10) \citet{diasetal2021w}; 11) \citet{piattietal2017a};
12) \citet{piatti2023a}.}
\end{table*}

\section{Data analysis}

In order to compile a list of open clusters included in the MWSC catalog
with ages larger than 1 Gyr, not detected by \citet{andersetal2021}, we 
compared both catalogs using
the IRAF\footnote{https://iraf-community.github.io/} {\sc ttools.tdiffer} task,
which creates an output table that includes only the rows that differ between two 
input tables. 
We note that the \citet{andersetal2021}'s sample consists of open clusters
located closer that 2.0 kpc from the Sun, so that we constrained our analysis to
those clusters. As a comparison variable, we employed
the clusters' names, and from the resulting list of clusters we selected those 
older than 1 Gyr, which turned out to be 136 open clusters. 
Before cross-matching the tables, which are very well manageable, we 
took care of spaces, underscores, different abbreviations, multiple names, etc, so 
that, we are confident that cross-matching names was in this case more secure than 
using coordinates, etc. 

We then searched the literature seeking for detailed studies independent from  
\citet{kharchenkoetal2013} and  \citet{andersetal2021}, focused on 
these 136 open clusters. This is to guarantee a third party analysis that
give us an independent assessment on the selected open clusters.
We found 17 open clusters that comply with that precepts. They are
listed in Table~\ref{tab1} with the respective references. As can be seen,
the detail studied open clusters represent $\sim$ 13 per cent
of the whole sample (136). This shows that most of the 136 selected open clusters
have not been studied in detail other than 
by \citet{kharchenkoetal2013} and/or \citet{andersetal2021},
which justifies to be embarked in this work.

Table~\ref{tab1} shows that previous detailed works on some open clusters
have confirmed their physical nature. We rely on these works as a support for 
the existence of these objects as real open clusters older than 1 Gyr. From
this point of view, we assume that the lack of detection of them 
by \citet{andersetal2021} could be caused by some of the reasons described in
\citet{cantatgaudinetal2018}. Nevertheless, there is one object,
ESO~436-02, which was also discarded by \citet{piattietal2017a}
as a genuine star aggregate (see Table~\ref{tab1}). We think that if more detailed 
studies of MWSC open clustesr were carried out, some of them could
be confirmed as real physical systems. Only 5 out of 10 old open clusters
in Table~\ref{tab1} (Bica~6, ESO~425-15, ESO~447-29, ESO~552-05, NGC~7036) are 
located within 2.0 kpc from the Sun (see references in Table~\ref{tab1}), so 
that they represent nearly a 2 per cent 
increase in the \citet{andersetal2021}'s old open cluster sample. The remaining
old open clusters in Table~\ref{tab1} have heliocentric distances from 2.3 up 
to 6.3 kpc, with and average of 4.0 kpc. We note that their heliocentric 
distances in the MWSC catalog are smaller than 2.0 kpc.

The age estimates of open clusters older than 1 Gyr in Table~\ref{tab1},
although derived from different studies, are in a general agreement
with those of the MWSC catalog. We obtained an average value and dispersion of
log(age)$_{\rm our}$ - log(age)$_{\rm MWSC}$) = 0.02$\pm$0.21. However, we 
found that 6 MWSC old open clusters resulted to be younger from detailed
independent works. We think that this discrepancy arises from constraints
in the star field decontamination procedure of  open cluster color-magnitude 
diagrams (CMDs) \citep{kharchenkoetal2013}, which could mislead the fitting of 
theoretical isochrones. This is also the case of the not confirmed open cluster 
ESO~436-02.

\begin{figure*}
\includegraphics[width=\textwidth]{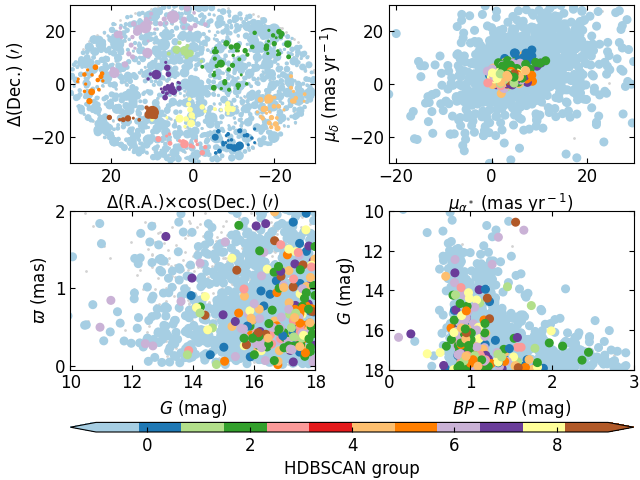}
\caption{Diagnostic diagrams from the HDBSCAN analysis for stars in the field of NGC~1520,
for which HDBSCAN identified  an unphysical stellar system.
Colored points correspond to nine different HDBSCAN groups of stars in the 5D-phase space.}
\label{fig1}
\end{figure*}

\subsection{HDBSCAN analysis}

From the above analysis, there are still 119 open clusters not included
in the compilation of \citet{andersetal2021}, for which the MWSC catalog provides 
age estimates larger than 1 Gyr. As far as we are aware, these objects do not
have in the literature any independent studies. Because of the confirmation of these 
objects as old open clusters is important in order to have a comprehensive knowledge 
of the older end of the Milky Way open cluster age distribution, we decided
to analyze them using independent data sets and analysis methods.

We retrieved from the {\it Gaia} DR3 \citep{gaiaetal2016,gaiaetal2022b} database
R.A. and Dec. coordinates, parallaxes ($\varpi$), proper motions
in R.A. and Dec. (pmra, pmdec), with their associated uncertainties, and $G$, $BP$, and 
$RP$ magnitudes of stars located inside circles with a radius of 30 arcmin from the 
centers of these 119 open clusters. We filtered the data 
following the recommendations described by \citet{cantatgaudinetal2018} and imposed the 
following cuts: $G$ $<$ 18 mag, and $|$pmra$|$,$|$pmdec$|$  $<$ 30
mas/yr \citep[see][]{hr2021}. 

A real open cluster is featured by being a spatial stellar overdensity, 
composed of stars located at a nearly same distance from the Sun and sharing
a mean motion. These conditions can be used by any clustering search engine to identify
open clusters in the {\it Gaia} DR3 database. 
We used the recommended HDBSCAN 
\citep[Hierarchical Density-Based Spatial Clustering of Applications with Noise,][]{campelloetal2013}
Gaussian mixture model technique \citep{hr2021} to search for overdensities in the 5D-phase space 
defined by R.A., Dec., $\varpi$, pmra, and pmdec. The \texttt{min$\_$cluster$\_$size} parameter 
was varied between 4 and 15 dex in steps of 1 dex, and from each output we built diagnostic plots 
as illustrated in Fig.~\ref{fig1},  where HDBSCAN identified stellar groups.
We colored the points according to the \texttt{clusterer.labels$\_$} parameter, which labels the
different identified groups of stars in the 5D-phase space. As can be seen, not only a group of
stars is identified close to the ($\Delta$(R.A.)$\times$cos(Dec.),$\Delta$(Dec.)) = (0,0)
(centered on the object), but also across the searched field. 

\begin{figure*}
\includegraphics[width=\textwidth]{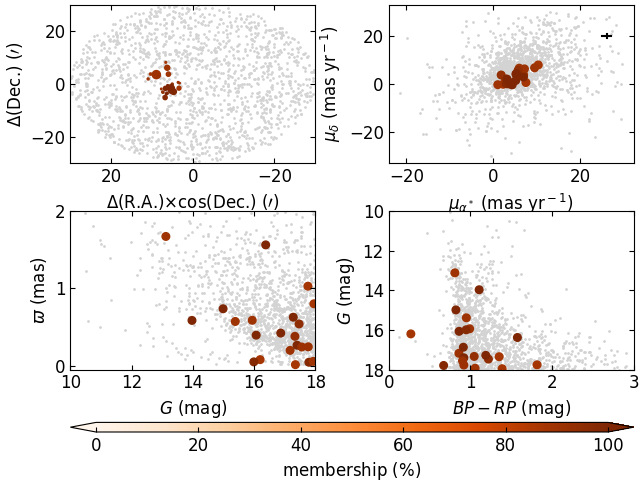}
\caption{Diagnostic diagrams for a selected group of stars in the field of NGC~1520
chosen from Fig.~\ref{fig1}. Colored points represent different membership probabilities. Grey
dots represent the whole {\it Gaia} DR3 data sets used.  The object was discarded
as an open cluster because it does not show the expected small dispersion in the Vector Point diagram
($\Delta$(pmra),$\Delta$(pmdec)) $\sim$  (1 mas/yr, 1 mas/yr), (see top-right panel), neither a relative constant trend
with $\varpi$ in the $\varpi$ versus $G$ diagram, nor a CMD with star sequences that suggests 
the presence of an old open cluster. NGC~1520 was removed from the NGC catalog by 
\citet{sulenticetal1973}, as pointed out by \citet{cga2020}, who also concluded that it is not an open cluster.  Symbol size in the top-left panel is proportional to the star brightness.}
\label{fig2}
\end{figure*}

The number of groups and the stars
included  in them can vary with the \texttt{min$\_$cluster$\_$size} parameter. Therefore, we
visually inspected the four panels of Fig.~\ref{fig1} 
looking for the optimum \texttt{min$\_$cluster$\_$size} value towards which the
\texttt{clusterer.labels$\_$} values remain constant and with similar
star distributions.  Each \texttt{clusterer.labels$\_$} value corresponds to a particular group of stars in Fig,~\ref{fig1}. Once we chose a group of stars, and hence its respective  
\texttt{clusterer.labels$\_$} value, we built Fig.~\ref{fig2} for all its stars, 
which shows the distribution of the selected 5D-phase space clustered stars in four different
plots. HDBSCAN provides also the membership probability of each star to the corresponding group. 
For the sake of the reader, Fig.~\ref{fig2} illustrates an example of a group of stars not confirmed 
as an open cluster (see Table~\ref{tab2}). If the chosen group of stars  shows the expected small 
dispersion in the Vector Point diagram
($\Delta$(pmra),$\Delta$(pmdec)) $\sim$  (1 mas/yr, 1 mas/yr) \citep{hr2021}; a relative constant trend
with $\varpi$ in the $\varpi$ versus $G$ diagram; and a CMD with star sequences that suggests the
presence of an old open cluster, we selected that object as a possible candidate for a further detailed 
analysis.  Figs.~\ref{fig3} and \ref{figa} show the number of stars for each candidate
cluster, colored according to their membership probabilities.

Table~\ref{tab2} lists the names of the objects that could not be confirmed as old open clusters, because
their diagnostic plots do not satisfy the above requirements. We are aware of the uncertainties
in the {\it Gaia} DR3 data sets, particularly in the 14 $\le$ $G$ (mag) $\le$ 18 range used, namely: 
$\sigma$($\varpi$) $\le$ 0.2 mas; $\sigma$(pmra, pmdec) $\le$ 0.2 mas/yr; $\sigma$($G$) $\le$ 0.01 mag,
$\sigma$(BP, RP) $\le$ 0.02 mag \citep{gaiaetal2022b,gaiaetal2022c}. These uncertainties do not
affect the assessments made on the diagnostic plots (see, e.g., Fig.~\ref{fig2}). As an exercise, 
we took into account  the parallax uncertainties (parallax is the most uncertain parameter) for  all the examined open clusters to extensively test HDBSCAN by using parallax 
values generated randomly from a normal distribution using the respective mean values and associated 
errors. For the individual executions we repeated the above analysis and recovered
\texttt{min$\_$cluster$\_$size} and \texttt{clusterer.labels$\_$} values that led us to conclude on the
same object status found previously. Table~2 lists 110 objects, which represent nearly 80$\%$ of the open
clusters older than 1 Gyr catalogued by \citet{kharchenkoetal2013} that were not included in the
compilation by \citet{andersetal2021}. 

\begin{table*}
\caption{Open clusters in the MWSC catalog not
included in the \\\citet{andersetal2021}'s compilation and not confirmed 
as older than 1 Gyr in \\this work by running HDBSCAN as a diagnostic test.}
\label{tab2}
\begin{tabular}{lllll}\hline

ASCC~48			&ESO~456-72	&FSR~0700	&FSR~1489	&NGC~1641\\
ASCC~57			&ESO~464-09	&FSR~0717	&FSR~1554	&NGC~2132\\
C1331-622		&ESO~502-19	&FSR~0724	&FSR~1574	& NGC~2143\\
Carraro~1			&ESO~570-12	&FSR~0733	&FSR~1577	& NGC~2189\\
Collinder~21		&FSR~0014	&FSR~0817	&FSR~1579	&NGC~2240\\
Collinder~196		&FSR~0050	&FSR~0820	&FSR~1607	&NGC~2306\\
Dol-Dzim~2		&FSR~0085	&FSR~0821	&FSR~1631	&NGC~2319\\
Dol-Dzim~6		&FSR~0091	&FSR~0843	&FSR~1652	&NGC~2348\\
Dolidze~17		&FSR~0110	&FSR~0886	&FSR~1678	&NGC~3909\\
Dolidze~38		&FSR~0112	&FSR~1050	&FSR~1682	&NGC~6738\\
Dolidze~50		&FSR~0117	&FSR~1066	&FSR~1685	&NGC~6856\\
Dutra-Bica~45		&FSR~0120	&FSR~1078	&FSR~1692	&NGC~6991A\\
ESO~008-06		&FSR~0128	&FSR~1197	&FSR~1695	& NGC~7050\\
ESO~043-13		&FSR~0265	&FSR~1204	&FSR~1705	&NGC~7084\\
ESO~129-19		&FSR~0266	&FSR~1293	&FSR~1719	&NGC~7772\\
ESO~132-14		&FSR~0307	&FSR~1387	&FSR~1729	&PTB~9\\
ESO~245-09		&FSR~0351	&FSR~1416	&FSR~1773	&Ruprecht~22\\
ESO~282-26		&FSR~0596	&FSR~1434	&IC~1023		&Ruprecht~103\\
ESO~329-02		&FSR~0689	&FSR~1442	&Latham~1	&Ruprecht~131\\
ESO~425-06		&FSR~0691	&FSR~1459	&Loden~894	&Ruprecht~139\\
ESO~435-33		&FSR~0692	&FSR~1465	&NGC~1252	&Ruprecht~146\\
ESO~442-04		&FSR~0695	&FSR~1467	&NGC~1520	&Stock~11	\\\hline
\end{tabular}
\end{table*}

There are still 9 remaining objects from the HDBSCAN analysis whose
diagnostic plots hint at the possibility of being open clusters older 
than 1 Gyr. When running HDBSCAN, proper motions resulted to be
the variables with much clearer clustering; in most of the cases
with points' dispersion smaller than a couple of mas/yr. Moreover, HDBSCAN
identified only one group of points in the Vector Point diagrams for each
one of these objects. All the stars considered resulted to be spatially
distributed in different groups (see Fig.~\ref{fig1}), which suggests that
proper motions alone cannot be used as a driven parameter to detect
open clusters. \citet{jaehnigetal2021} recently based their discovery
of new 11 open clusters on Vector Point diagrams' overdensities using
{\it Gaia} DR2 data sets. The 11 new objects exhibit CMD star sequences
resembling those of open clusters. However, \citet{piattietal2023}
showed that the dispersion of their fundamental properties (age, distance, 
reddening, metallicity) turned out to be much larger than those usually 
obtained for open clusters. Indeed, they resemble those of ages and metallicities 
of composite star field populations, or possibly sparse groups of stars. 
This result prevent us of using proper motions as a main drivers for
identifying real open clusters.

Parallaxes resulted the less clustered variable while running HDBSCAN, except
in the cases of the stars of the 9 aforementioned open clusters. Nevertheless, 
their $\varpi$ versus $G$ diagrams need some additional cleaning of interlopers; 
mainly stars located in the line of sight towards the open clusters with 
proper motions similar to that of cluster members. 

\subsection{Color-magnitude diagram analysis}

\begin{figure*}
\includegraphics[width=\textwidth]{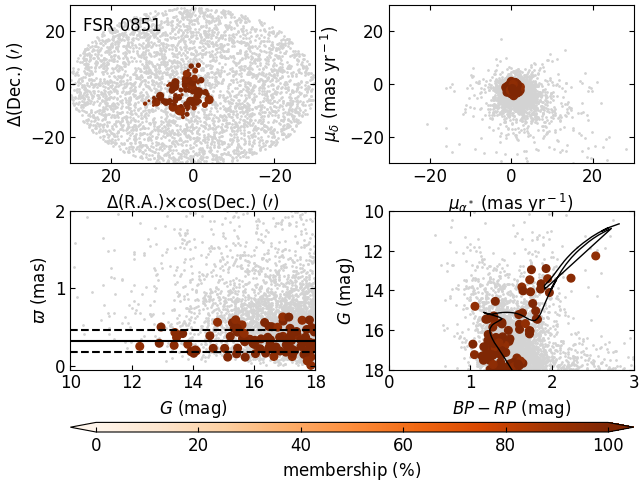}
\caption{Same as Fig.~\ref{fig2} for  FSR~0851. The $\varpi$ vs $G$ plot
shows the mean and standard deviation of $\varpi$ drawn with solid and
dashed lines respectively. The CMD shows the best fitted isochrone
superimposed.}
\label{fig3}
\end{figure*}

The CMDs of these 9 open clusters also show the presence
of field stars. We used Figs.~\ref{fig3} and \ref{figa} to further analyze these objects.
Firstly, we derived the mean and dispersion of the open cluster parallaxes, which
are represented by solid and dashed lines in the $\varpi$ versus $G$ diagrams (bottom-right
panels), respectively. From these parallaxes and the open clusters' central coordinates, we
obtained mean reddening and dispersion using different Milky Way reddening map 
models provided through the GALExtin\footnote{http://www.galextin.org/} interface 
\citep{amoresetal2021}. 

In order to decontaminate field stars in the cluster CMDs we relied 
on the procedure devised by  \citet{pb12}, which has been shown to produce 
cleaned cluster CMDs \citep[e.g.,][and references therein]{pl2022,piatti2022}. 
The method consists in comparing the cluster CMD with the CMD of a reference star field
located adjacent to the cluster region and subtracting from the cluster CMD the closest 
stars to those
in the respective reference star field CMD. Fig.~\ref{fig3} illustrates the
cleaned diagnostic plots for FSR~0851. The field star cleaned diagnostic plots for
the remaining 8 open clusters are depicted in the Appendix (see Fig.~\ref{figa}).

We used the independent measures of distance and reddening to 
guide the Automated Stellar Cluster Analysis
code \citep[\texttt{ASteCA,}][]{pvp15} in deriving the clusters'  ages and metallicities.
\texttt{ASteCA} explores the parameter space  of 
synthetic CMDs through the minimization of the likelihood function defined by 
\citet[][the Poisson likelihood ratio (eq. 10)]{tremmeletal2013} using a parallel tempering 
Bayesian MCMC algorithm, and the optimal binning \citet{knuth2018}'s method.
To generate the synthetic CMDs, \texttt{ASteCA} uses the theoretical isochrones computed by
\citet[][PARSEC v1.2S\footnote{http://stev.oapd.inaf.it/cgi-bin/cmd}]{betal12}, the initial 
mass function of \citet{kroupa02} and cluster masses in the range 100-5000M$_\odot$, whereas 
binary fractions are allowed in the range 0.0-0.5 with a minimum mass ratio of 0.5.
Table~\ref{tab3} lists the resulting cluster astrophysical properties, while Figs.~\ref{fig3}
and \ref{figa} show the respective theoretical isochrones superimposed onto the cleaned 
cluster CMDs.

The  9 analyzed objects resulted to be old open clusters; the mean difference between the 
present values and the ages listed in the MWSC catalog being 
log(age)$_{\rm our}$ - log(age)$_{\rm MWSC}$) = -0.14 $\pm$ 0.12. Albeit they are old
open clusters, their derived
heliocentric distances resulted to be larger than 2.0 kpc, so that they
cannot be added to the present comparison between \citet{andersetal2021}'s compilation and
the MWSC catalog. Heliocentric distances larger than 2.0 were also derived for half of the
old open clusters listed in Table~\ref{tab1} with detailed independent studies 
(MWSC distances $<$ 2.0 kpc).  
By inspecting the cluster CMDs with superimposed theoretical isochrones shifted 
using the MWSC distances, we found that they match very well the sequences of field stars.
This means that contamination of field stars is at some level present in the cluster CMDs
used by \citet{kharchenkoetal2013}. Perhaps, this contamination  may also explain the
recognition of many objects as open clusters that {\it Gaia} data combined with
machine learning techniques could not be able to recover them (see, e.g. Table~\ref{tab2}).

\begin{table*}
\caption{Derived properties for the studied open clusters.}
\label{tab3}
\begin{tabular}{lccccccc}\hline
Name        & $E(B-V)$ & $(m-M)_0$ & d      & [Fe/H] & log(age /yr) & log(age /yr)$_{\rm MWSC}$ & d$_{\rm MWSC}$  \\
            &   (mag)  &   (mag)   & (kpc)  & (dex) &   & &(kpc) \\\hline
ESO~427-32  & 0.21$\pm$0.05 & 13.90$\pm$0.24 &6.0$\pm$0.4 & 0.05$\pm$0.16 & 9.10$\pm$0.06 & 9.11 & 1.79\\
FSR~0121    & 0.32$\pm$0.06 & 13.49$\pm$0.23 &5.0$\pm$0.3 &-0.22$\pm$0.17 & 9.60$\pm$0.14 & 9.45 & 1.52\\
FSR~0145    & 0.26$\pm$0.04 & 13.31$\pm$0.25 &4.5$\pm$0.3 &-0.19$\pm$0.20 & 9.50$\pm$0.12 & 9.19 & 1.61\\
FSR~0616    & 0.27$\pm$0.03 & 13.40$\pm$0.30 &4.8$\pm$0.4 &-0.49$\pm$0.24 & 9.35$\pm$0.07 & 9.23 & 1.53\\
FSR~0690    & 0.47$\pm$0.04 & 13.00$\pm$0.19 &4.0$\pm$0.2 &-0.03$\pm$0.15 & 9.30$\pm$0.13 & 9.00 & 1.71 \\
FSR~0851    & 0.69$\pm$0.06 & 12.10$\pm$0.30 &2.6$\pm$0.2 &-0.44$\pm$0.20 & 9.30$\pm$0.11 & 9.15 & 1.09 \\
FSR~1566    & 0.16$\pm$0.05 & 12.50$\pm$0.20 &3.2$\pm$0.2 & 0.19$\pm$0.19 & 9.55$\pm$0.09 & 9.43 & 1.54\\
NGC~2395    & 0.21$\pm$0.05 & 12.73$\pm$0.28 &3.5$\pm$0.4 & 0.03$\pm$0.13 & 9.20$\pm$0.08 & 9.28 & 1.21\\
Ruprecht~3  & 0.13$\pm$0.05 & 13.50$\pm$0.22 &5.0$\pm$0.3 &-0.04$\pm$0.14 & 9.30$\pm$0.09 & 9.10 & 1.26\\\hline
\end{tabular}
\noindent Note: log(age /yr)$_{\rm MWSC}$ and d$_{\rm MWSC}$ are the age and the heliocentric distance in the 
MWSC catalog, respectively.
\end{table*}

\section{Discussion and concluding remarks}

The present detailed analysis of 136 open clusters included in the MWSC catalog,
carried out from a dedicated
HDBSCAN clustering search and a powerful technique for the decontamination 
of field stars in the cluster CMDs, shows that they all are not older than 1 Gyr.
We found that 19 out of the 136 objects analyzed are real old open clusters; the
remaining ones (117) being younger open clusters  (6) or not confirmed physical
systems  (111). Five out of the 19 confirmed old open clusters are located inside a circle 
of 2 kpc from the Sun. They represent an increase
of $\sim$ 2$\%$ in the compilation of clusters older than 1 Gyr by 
\citet{andersetal2021}. This outcome shows that detailed studies
are necessary to disentangling the real nature of catalogued open clusters,
as well as that the MWSC catalog contains a large percentage  (29$\%$) of
non-real stellar aggregates among those with assigned ages larger
than 1 Gyr. In brief, the MWSC catalog  supersedes by those built from {\it Gaia}
data.

Nevertheless, we also note that \citet{hr2023} showed that there is a number of limitations 
of HDBSCAN and  in {\it Gaia} data, and differences in the quality cuts and definitions of an 
open cluster that can lead us to be unable to detect some open clusters. Besides, a 
dedicated procedure to clean the field star contamination in cluster CMDs is also needed.
Indeed, while performing an all-sky census of open clusters using the {\it Gaia} 
DR3 database and HDBSCAN, they did not detect 1152 open cluster in the MWSC catalog
and built an open cluster age function which, for ages larger than 1 Gyr, includes
even less clusters than \citet{andersetal2021}. They also show that the open cluster age
distribution constructed by \citet{kounkeletal2020} from {\it Gaia} data for that age range is in excellent 
agreement with that by \citet{kharchenkoetal2013}  (see their figure~11), which seems
somehow paradoxical. Furthermore, other studies based on the {\it Gaia} database have
even found more old open clusters  \citep{heetal2022a,heetal2022b,heetal2023} and \citet{haoetal2022}, which in turn, resulted in open cluster age distributions
with an excess of old open clusters with respect to that built by \citet[][see their figure~15]{hr2023}.
 
 We performed a search of the 9 open clusters confirmed in this work (see Table~\ref{tab3}) in the
 catalogs built by \citet{kounkeletal2020}, \citet{heetal2022a}, \citet{heetal2022b}, \citet{heetal2023},
 and \citet{hr2023} and found none of them. This outcome reinforces the verdict that the 
 different clustering search methods can produce different outcomes.

Previous discrepancies about the completeness of the known old open cluster population, and hence
of its age distribution function, are also found in the literature. For instance, 
\citet{kharchenkoetal2013} claimed that the
MWSC catalog is almost complete out to 1.8 kpc  from the Sun, except possibly for
clusters older than 1 Gyr, a result that was used as a support by \citet{joshietal2016} in their
analysis of the Galactic structure and to conclude on different relationships between the cluster's
mass, age, and diameter. On the other hand, 
\citet{andersetal2021} estimated a completeness of $\sim$ 88$\%$
of their old open cluster sample based on open cluster recovery experiments. They performed 
those experiments using as a reference the catalog of \citet{castroginardetal2020},
which was also built from {\it Gaia} DR2 data. Curiously, both \citet{kharchenkoetal2013} 
and \citet{andersetal2021} mentioned that their cluster samples are almost complete, although
the former include $\sim$ 50$\%$ more old open clusters than the latter.
The above results  show that our knowledge of the old end of the open cluster population 
is far from being complete, so that definitive conclusions on their properties 
could be risky to draw at the present completeness of the old open cluster
population.

We think that some of the present constraints of building a statistically complete
old open cluster age distribution function in the solar neighborhood could be mitigated 
from deeper imaging surveys,
which could help identifying uncovered old open clusters. From currently available 
imaging surveys, recent works applying machine learning techniques 
\citep[see pros and cons of different methods in][]{hr2021} 
to identify open clusters, 
and particularly new discoveries \citep[see a summary compiled in Table~3 in][]{hr2023},
present mainly open clusters with relatively long Main Sequences ($>$ 6 mag long). 
However, distant old open clusters do not show long Main Sequences down 
to $G$ = 18 mag (see Figs~\ref{fig3} and \ref{figa}), which is the most common
limiting magnitude used when dealing with {\it Gaia} data. 

Summarizing, we did not find any remarkable difference between the population of 
open clusters older than 1 Gyr in \citet{andersetal2021} and in its counterpart 
in the MWSC catalog, although there are still doubts about their 
completeness.  As mentioned above. other recent searches for new 
open clusters have found more old objects using  also {\it Gaia} data, which poses the issue 
about the different performances of the devised detection procedures, including 
machine learning methods and cleaning of the field stars in their CMDs. Therefore, a
comprehensive solar neighborhood old open cluster age distribution function is still under 
construction, and will require much more effort focused, among others, on deeper imaging 
surveys. From this point of view, the claims by \citet{andersetal2021} about a remarkable
earlier drop of the old open cluster population should be considered to the light of the above
mentioned challenges. We think that the careful analysis carried out in this work 
sheds light into the advantages and disadvantages of different data sets and analysis
procedures. As far as we are aware, they have been highlighted in the literature unevenly.

\section*{Acknowledgements}
We thank the referee for the thorough reading of the manuscript and timely 
suggestions to improve it.
This work has made use of data from the European Space Agency (ESA) mission
{\it Gaia} (https://www.cosmos.esa.int/gaia), processed by the {\it Gaia}
Data Processing and Analysis Consortium (DPAC,
https://www.cosmos.esa.int/web/gaia/dpac/consortium). Funding for the DPAC
has been provided by national institutions, in particular the institutions
participating in the {\it Gaia} Multilateral Agreement.

This research made use of Astropy, a community-developed
core Python package for Astronomy.



\appendix

\section{HDBSCAN and CMD analyses}

In this Section we present the diagnostic diagrams of MWSC
open clusters older than 1 Gyr according to \citet{kharchenkoetal2013},
not included in the compilation by \citet{andersetal2021}, for which
we confirmed their older age (see Section 2.2).

\begin{figure*}
\includegraphics[width=\columnwidth]{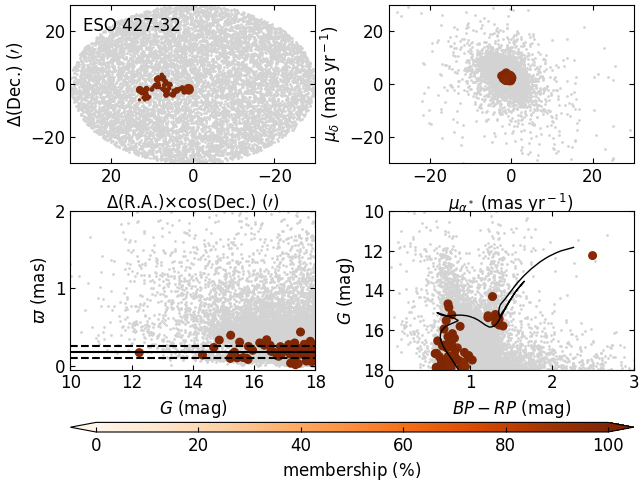}
\includegraphics[width=\columnwidth]{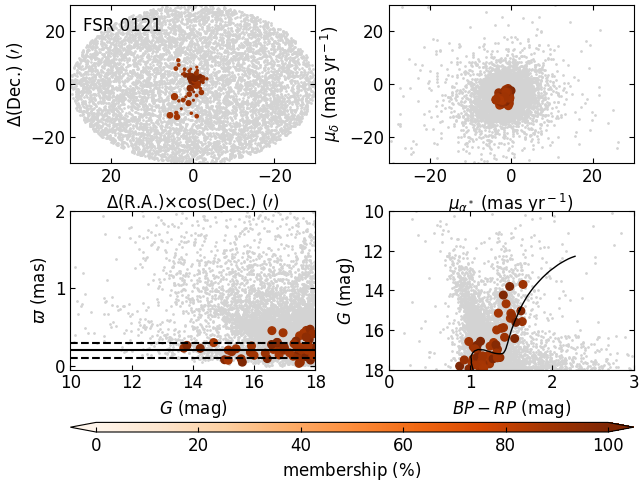}
\includegraphics[width=\columnwidth]{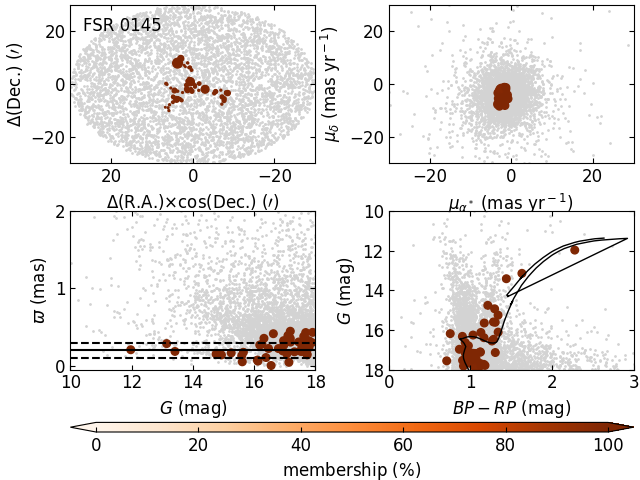}
\includegraphics[width=\columnwidth]{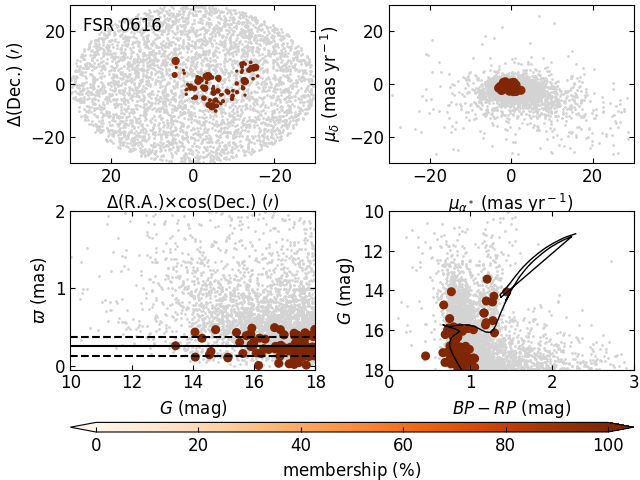}
\includegraphics[width=\columnwidth]{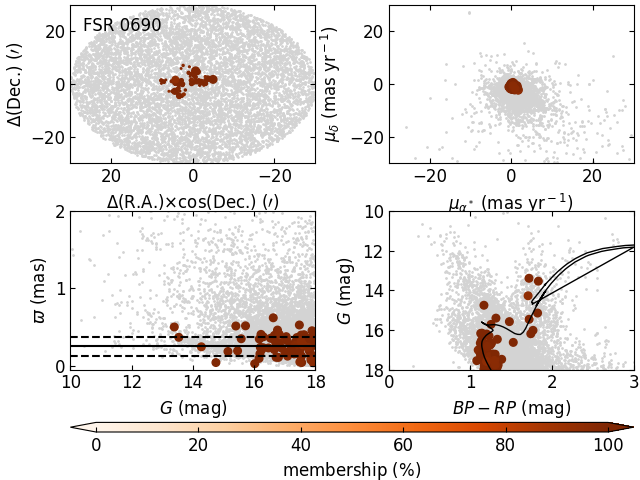}
\includegraphics[width=\columnwidth]{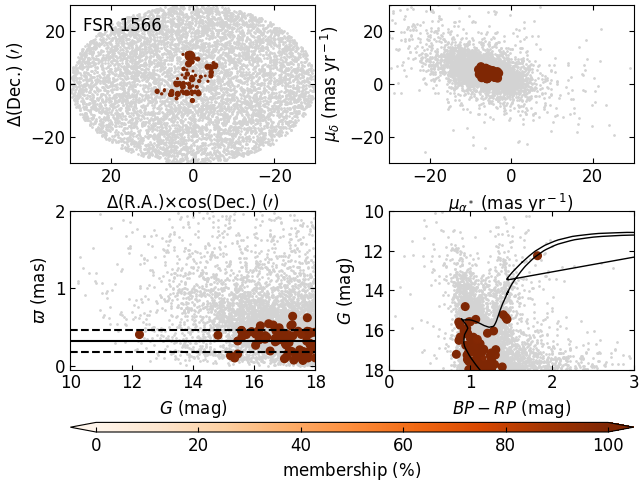}
\caption{Same as Fig.~\ref{fig3} for analyzed MWSC open clusters older than 1 Gyr. 
The $\varpi$ vs $G$ plot shows the mean and standard deviation of $\varpi$ drawn with 
solid and dashed lines respectively. The CMD shows the best fitted isochrone
superimposed.}
\label{figa}
\end{figure*}

\setcounter{figure}{0}
\begin{figure*}
\includegraphics[width=\columnwidth]{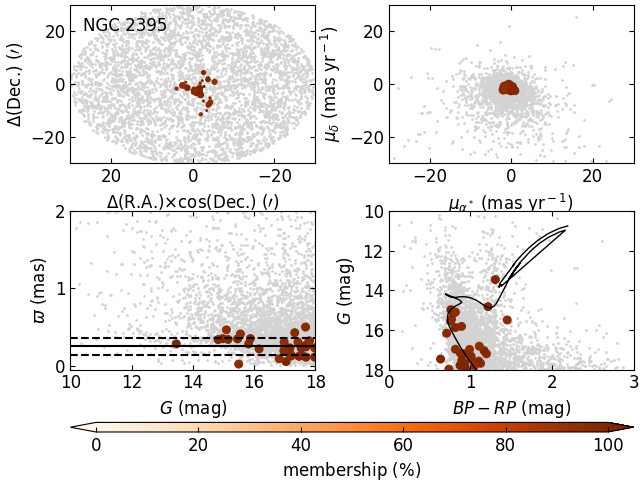}
\includegraphics[width=\columnwidth]{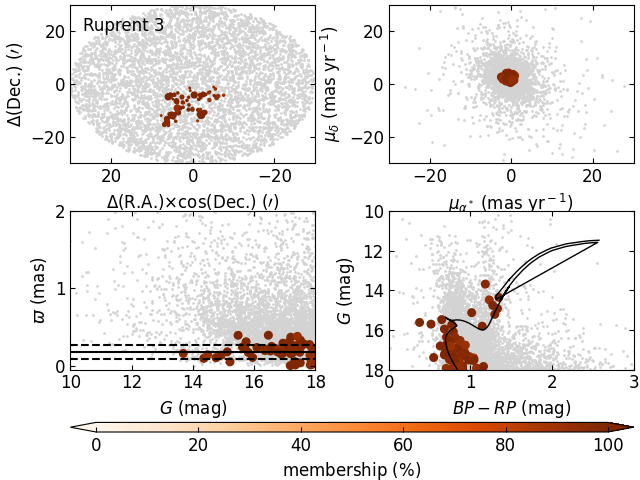}
\caption{continued.}
\end{figure*}

\end{document}